\documentclass[]{raa}            

\usepackage{graphicx,times}             
\input{epsf.sty}                        

\begin{document}

\title{Hydro-chemical study of the evolution of interstellar pre-biotic molecules during the collapse of molecular clouds}

   \volnopage{Vol.0 (200x) No.0, 000--000}      
   \setcounter{page}{1}          

   \author{Liton Majumdar
      \inst{1}
   \and Ankan Das
      \inst{1}
   \and Sandip K. Chakrabarti
      \inst{2,1}
   \and Sonali Chakrabarti
      \inst{3,1}
   }

   \institute{Indian Centre for Space Physics, Chalantika 43, Garia Station Road, 
             Kolkata- 700084, India; {\it ankan@csp.res.in, liton@csp.res.in}\\
         \and
              S. N. Bose National Centre for Basic Sciences, Salt Lake, 
              Kolkata- 700098, India; {\it chakraba@bose.res.in}\\
         \and
             Maharaja Manindra Chandra College, 20 Ramakanto Bose Street, 
             Kolkata- 700003, India; {\it sonali@csp.res.in}\\
     }

   \date{Received~~2009 month day; accepted~~2009~~month day}

\abstract{One of the stumbling blocks for studying the evolution of interstellar 
molecules is the lack of adequate knowledge of the rate co-efficients of 
various reactions which take place in the Interstellar medium and molecular clouds.
Some of the theoretical models of rate coefficients  do exist 
in the literature for computing abundances of the complex pre-biotic molecules.
So far these have been used to study the abundances of these molecules in space. 
However, in order to obtain more accurate final compositions in these media, 
we find out the rate coefficients for the formation of some of the most 
important interstellar pre-biotic molecules by using quantum chemical
theory. We use these rates inside our hydro-chemical model to
find out the chemical evolution and the final abundances of the pre-biotic species 
during the collapsing phase of a proto-star. We find that a significant 
amount of various pre-biotic molecules could be produced during the
collapsing phase of a proto-star. We study extensively the formation 
these molecules via successive neutral-neutral and radical-radical/radical-molecular reactions. 
We present the time evolution of the chemical species with an emphasis on  
how the production of these molecules varies with the depth of a cloud. We compare
the formation of adenine in the interstellar space using our rate-coefficients 
and using those obtained from the existing theoretical models. 
Formation routes of the pre-biotic molecules are found to be highly dependent on the
abundances of the reactive species and the rate coefficients involved in the reactions.
Presence of grains strongly affect the abundances of the gas phase species.
We also carry out a comparative study between different pathways available for the synthesis of
adenine, alanine, glycine and other molecules considered in our network.
Despite the huge abundances of the neutral reactive species, production of adenine 
is found to be highly dominated by the radical-radical/radical-molecular reaction pathways.
If all the reactions considered here are contributing for the production of alanine and 
glycine, then neutral-neutral \& radical-radical/radical-molecular pathways both are found to have
significant contribution for the production of alanine, whereas 
radical-radical/radical-molecular pathways plays a major role in case of glycine production.
}

   \authorrunning{Liton Majumdar, Ankan Das, Sandip K. Chakrabarti, Sonali Chakrabarti }            
   \titlerunning{Hydro-chemical modeling of the formation of interstellar pre-biotic molecules }           

   \maketitle

%
%
\section{Introduction}           
According to the CDMS catalog (http://www.astro.uni-koeln.de/cdms/molecules)
approximately 165 molecules have been detected in the interstellar medium or circumstellar shells.
The significance of the interstellar dust towards the formation of complex interstellar
molecules is well recognized after the discovery of more than 20 
molecules in the interstellar ice, especially around the star
forming regions (Boogert \& Ehrenfreund 2004, Gibb et al., 2004 etc.).
Increasing experimental evidences also point to the importance
of surface processes (Ioppolo et al., 2008, Oberg et al., 2009 etc.) and suggest
to incorporate the grain surface processes extensively.
Indeed, a number of attempts were made to appropriately model the interstellar grain 
chemistry over the years (e.g., Das et al., 2006ab; Das et al. 2008b; Das et al., 
2010; Cuppen et al., 2007; Cuppen et al., 2010). 

At the same time, an understanding of the origin of life is a long standing problem and 
it is curious to know whether various pre-biotic molecules could also be formed in the Inter-stellar Medium. 
It was suggested that during the collapsing phase of a proto-star, some basic elements of life such as amino acids and some of the
bases of DNA, could be produced (Chakrabarti \& Chakrabarti 2000ab; hereafter CC00a and CC00b
respectively). 
 It is expected that there might be some good connection of nitrile chemistry with the prebiotic
chemistry. Motivated by detections of nitriles in Titans atmosphere, cometary comae, and the
interstellar medium, Hudson \& Moore (2004) reported laboratory investigations of low-temperature
chemistry of acetonitrile, propionitrile, acrylonitrile, cyanoacetylene, and cyanogen. They performed
a few experiments on different nitriles to produce low-temperature photo and radiation products
of nitriles.
Blagojevic et al., (2003) reported their experiment for the formation of gas phase glycine and
$\beta$-alanine, as well as acetic and propanoic acid, from smaller molecules in the
interstellar environments. The effects of the fluxes of charged particles and 
electromagnetic radiation also play an important role in the
chemical composition of the interstellar clouds (Das et al., 2011). 
Complex pre-biotic molecules might thus be formed due to very
complex and rich chemical processes inside the molecular cloud. The production of amino acids,
nucleotides, carbohydrates and other basic compounds are possibly starting from the 
molecules like HCN, cyno compounds, aldehyde,
and ketones (Orgel 2004; Abelson 1966), which could lead to the origin of life 
in the primitive earth conditions. Using an effective (gas-grain) reaction rate, CC00ab proposed the oligomerization of HCN 
to produce some adenine and other complex molecules.

In the present paper, we perform  quantum chemical calculations to find out
the rate coefficients for the formation of several pre-biotic molecules such as, adenine, alanine, glycine, glycolic acid and lactic acid. Obtained rate
coefficients from the quantum chemical calculations are directly implemented 
into our hydro-chemical code to find out the
chemical evolution during the collapsing phase of a proto-star. 

The plan of this paper is following. In Section 2, the models and the
computational details are discussed. Different computational
results are presented in Section 3, and finally in Section 4, we draw our conclusions.

\begin{table*}
\scriptsize{
\centering
\addtolength{\tabcolsep}{-4pt}
\caption{Estimated rate co-efficients for the formation of various pre- biotic molecules
in the gas phase based on B3LYP/6-31G** calculations and Bates (1983) semi-empirical formula}
\vspace{1cm}
\begin{tabular}{|c|c|c|c|c|}
\hline
  {\bf Species}&{\bf Neutral-Neutral}&{\bf Magnitude of
 }&{\bf Radical-radical/ }
&{\bf Magnitude of } \\
   &{\bf reaction pathways}&{\bf association energies (eV)/}&{\bf radical-molecular}&
{\bf association energies (eV)/} \\
   &&{\bf Rate coefficients }&{\bf reaction pathways}&{\bf Rate coefficients } \\
   &&{\bf (cm$^3$ s$^{-1}$)}&&{\bf (cm$^3$ s$^{-1}$)} \\
\hline
\hline
&&&&\\
&(A) HCN+HCN$\rightarrow$CH(NH)CN& 0.14 / 8.38 x10$^{-20}$&(a) HCCN+HCN$\rightarrow$Molecule 1& 0.6443 / 2.13 x10$^{-17}$\\
&(B) CH(NH)CN+HCN$\rightarrow$NH$_2$CH(CN)$_2$& 2.13 / 3.43 x10$^{-12}$&(b) Molecule 1 +H $\rightarrow$Molecule 2& 4.2362 / 7.96 x10$^{-9}$\\
Adenine&(C) NH$_2$CH(CN)$_2$+HCN$\rightarrow$& 1.12 / 3.30 x10$^{-15}$&(c) Molecule 2+NH$_2$CN
$\rightarrow$Molecule 3& 1.9019 / 6.24 x10$^{-12}$\\
&NH$_2$(CN)C=C(CN)NH$_2$&&(d) Molecule 3+CN $\rightarrow$Molecule 4& 3.6300 / 1.80 x10$^{-9}$\\
&(D) NH$_2$(CN)C=C(CN)NH$_2$+HCN$\rightarrow$& 2.31 / 3.99 x10$^{-10}$&(e) Molecule 4+H$\rightarrow$Molecule 5& 3.0380 / 8.71 x10$^{-9}$\\
&C$_5$H$_5$N$_5$&&(f) Molecule 5+CN$\rightarrow$C$_5$H$_5$N$_5$+HNC& 4.6738 / 1.89 x10$^{-9}$\\
&&&(g) Molecule 5+CN$\rightarrow$C$_5$H$_5$N$_5$+HCN& 5.1274/ 1.91 x10$^{-9}$\\
&&&&\\
\hline
&&&&\\
&(A) H$_2$CO+HCN$\rightarrow$C$_2$H$_3$ON& 2.40 / 1.23 x10$^{-12}$&(a) HCN+H$\rightarrow$HCNH& 0.64 / 5.56 x10$^{-18}$\\
Glycine&(B) C$_2$H$_3$ON+H$_2$O$\rightarrow$C$_2$H$_5$NO$_2$& 1.11 / 2.68 x10$^{-15}$&(b) HCNH+H$\rightarrow$CH$_2$NH
& 4.50 / 1.63 x10$^{-12}$\\
&&&(c) CH$_2$NH+H$\rightarrow$CH$_2$NH$_2$& 1.70 / 1.17 x10$^{-14}$\\
&&&(d) CO+OH$\rightarrow$COOH& 0.78/1.13 $\times 10^{-17}$\\
&&&(e) CH$_2$NH$_2$+COOH$\rightarrow$C$_2$H$_5$NO$_2$& 3.06 / 1.82 x10$^{-9}$\\
&&&&\\
\hline
&&&&\\
Alanine&(A) CH$_3$CHO+HCN$\rightarrow$C$_3$H$_5$ON& 3.58 / 1.84 x10$^{-9}$&(a) C$_2$H$_5$NO$_2$$\rightarrow$NH$_2$CHCOOH+H& 13.55 / 7.22 x10$^{-10}$\\
&(B) C$_3$H$_5$ON+H$_2$O$\rightarrow$C$_3$H$_7$NO$_2$& 1.33 / 3.09 x10$^{-14}$&(b) NH$_2$CHCOOH+CH$_3$$\rightarrow$C$_3$H$_7$NO$_2$& 2.06 / 2.30 x10$^{-11}$\\
&&&&\\
\hline
&&&&\\
Glycolic Acid&(A) C$_2$H$_5$NO$_2$+H$_2$O$\rightarrow$C$_2$H$_4$O$_3$+NH$_3$& 1.22/9.51 x10$^{-15}$&-&-\\
&&&&\\
\hline
&&&&\\
Lactic Acid&(A) C$_3$H$_7$NO$_2$+H$_2$O$\rightarrow$C$_3$H$_6$O$_3$+NH$_3$& 2.56/2.31 x10$^{-9}$&-&-\\
&&&&\\
\hline
\end{tabular}}
\end{table*}

\section{Methods and Computational Details}

\subsection{Quantum chemical calculations}
Density functional theory (DFT) is an efficient tool to explore the chemical parameters of a
species. We use the DFT formalism explicitly to find out the different chemical parameters
for the synthesis of interstellar pre-biotic molecules. The computations were performed using 
B3LYP functional (Becke, 1993)
with the 6-31G** basis set available in the Gaussian 03W package. Total energies, zero point
vibrational energy and electronic energies of all the species which are formed during the 
synthesis of different
interstellar pre-biotic molecules were calculated. Armed with these important chemical parameters, the desired
rate coefficients for any reactions at 30K have been calculated by using semi-empirical
relationship developed by Bates (1983), i.e.,
\begin{equation}
K = 1 \times 10^{-21} A_r (6E_0 + N -2)^{(3N-7)}/ (3N-7)!  cm^3 s^{-1}
\end{equation}
where, $E_0$ is the magnitude of association energy in eV, $A_r$ is the transition probability (in $s^{-1}$) of the
stabilizing transition (the numerical value of which may be taken to be 100 unless better information is
available) and N is the number of nuclei in the complex.

If the calculated rate coefficients from Eqn. 1 exceeds the limit set by the following equation
(Eqn. 2), then this limiting value should be adopted
\begin{equation}
K = 7.41 \times 10^{-10} \alpha^{1/2}(10/\mu)^{1/2} cm^3 s^{-1}
\end{equation}
where, $\alpha$ is the plorizibility in {A$^{\circ}$}$^3$, $\mu$ is the reduced mass of the reactants in $^{12}$C amu
scale as suggested by Bates (1983).
Though the rate expression reported by Bates (1983) is not a robust modern rate theory as it depends only
on the association energy not on the reaction barrier height and other relevant features of the reaction
surface, we are still considering this expression in order to get an estimation over the rate coefficient
for the formation of different interstellar pre- biotic molecules around the low temperature region of the molecular clouds.
Here, we have used different thermodynamical variables from Das et al., (2008a), 
which are belongs to 10K. Thus we are implementing Bates's (1983) relation at 10K also.

\begin {figure}
\vskip 0.6cm
\centering{
\includegraphics[height=6cm,width=8cm]{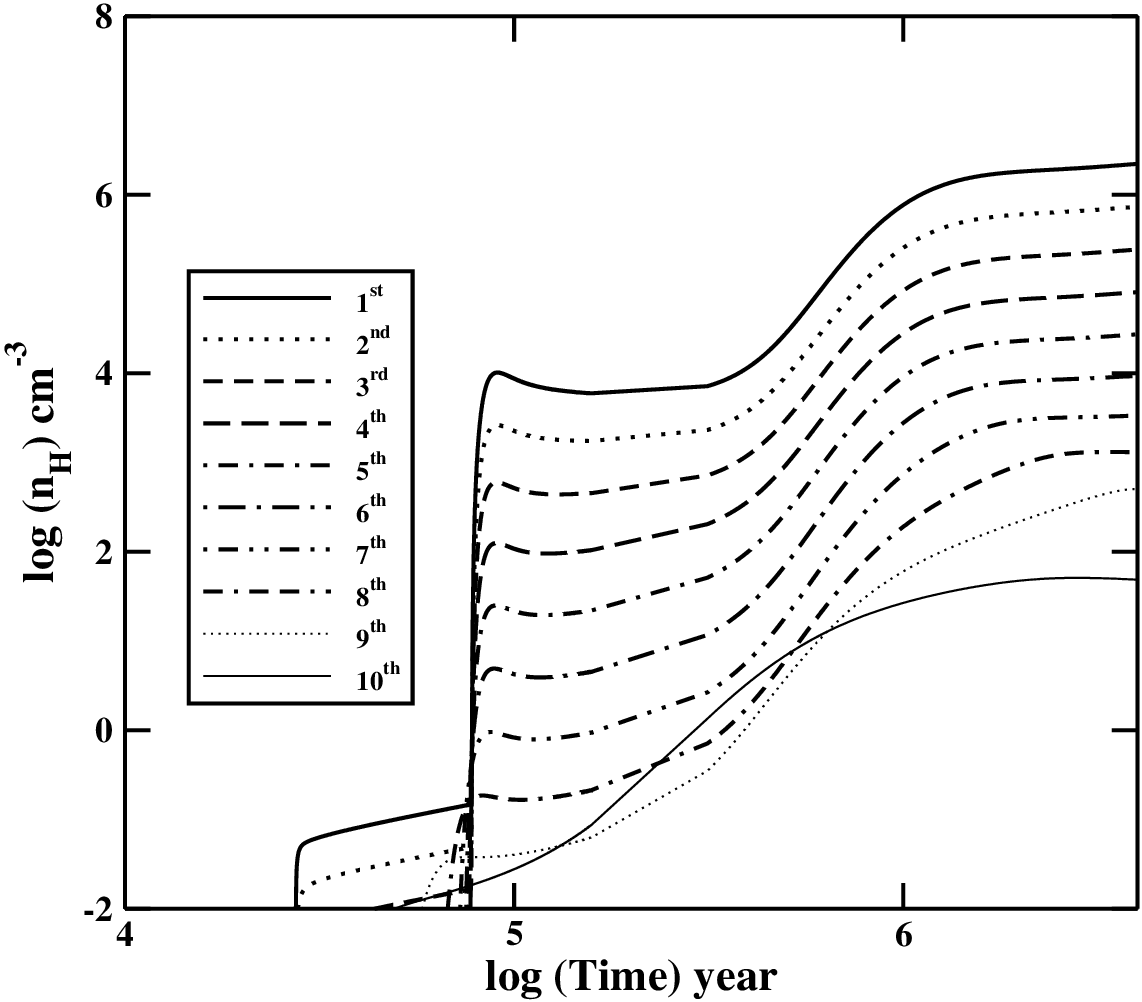}}
\caption{Time evolution of the number density is shown for the different region of the collapsing cloud.
}
\label{fig-1}
\end {figure}

\subsection{Hydro-chemical model}
Since the chemical evolution around the star forming region highly depends on the physical
properties of the ISM, it is necessary to have a realistic description of the collapsing cloud to have
accurate abundances of various interstellar species.
In the present paper, for the sake of simplicity, we assume the density distribution
as described in Das et al., (2008a) and Das et al., (2010). A spherically symmetric self-gravitating
interstellar cloud was considered in order to mimic the physical behavior during the star formation.
We assume that the outer boundary of the cloud is located at one parsec and the inner boundary is
located at $10^{-4}$ parsec. Matter entering through the inner boundary was assumed to be added
to the core mass. In Das et al., (2008a), the entire cloud was divided into $100$
logarithmically equally spaced grids in the radial direction. For the sake of simplicity in averaging,
we assume $10$ radial shells, so that one shell consists of $10$ radial grids. We took volume average
of the number densities for each of the $10$ grids to assign the number density of any particular shell.

We developed a large gas-grain network to explore the chemical evolution of a collapsing cloud.
For the gas-phase chemical network, we follow the UMIST 2006 data base. Formation of different
pre-biotic molecules are included in this gas phase chemical network. However, the rate coefficients for the
formation of several pre- biotic molecules are till date unknown. We calculate all these rate
coefficients and included in our gas phase network. For the grain surface reaction network,
we follow Hasegawa, Herbst \& Leung (1992), Cuppen et al., (2007), Das et al., (2010) and 
Das \& Chakrabarti (2011).
Our surface network consists of all the updated interaction barrier energies 
as mentioned in Das \& Chakrabarti, (2011) and
references therein. A detailed discussions about our gas-grain chemical model are already presented
in Das et al., (2011). Following the hydrodynamical model as discussed in Das et al., (2008a) 
and Das et al.,(2010), we consider the density distribution of a collapsing cloud as 
an input parameter of our gas-grain chemical code.

\section{Results and Discussion}

\subsection{Reaction pathways and rate coefficients}

There are several pathways by which a complex molecule can be formed. But since we are
focusing on the formation of these species around a star forming region, we need to
consider the most energetically economical route in which these species can be synthesized. Chemical
abundances of those species, which are identified as the precursor for the formation of pre-biotic molecules
such as HCN (Jiurys, 2006), HCCN (Guelin \& Cernicharo, 1991), NH$_2$CN (Turner et al., 1975),
CN (Fuente et al., 2005), C$_2$H$_5$, H$_2$CO, CH$_3$CHO (Woodall et al., 2007) etc. are dictating the reaction
pathways for the formation of the interstellar pre-biotic molecules. So it is essential to consider
the energy budget of any reaction along with the abundance of the reactive species.
There are several neutral-neutral as well as radical-radical/radical-molecular pathways
for the formation of these pre-biotic molecules. However, we consider only the most feasible pathways under the
interstellar circumstances. We concentrate mainly on the formation of adenine,
glycine, alanine, lactic acid and glycolic acid in the gas phase. The formation of these molecules
via neutral-neutral and radical-radical/radical-molecular pathways with their rate coefficients and 
association energies
(Energy difference in eV, with zero point corrected values)
are listed in Table 1. Formation of other pre-biotic molecules such as the glycolic acid, lactic acid via
only neutral-neutral reaction pathways (no evidences for the synthesis of these species  by the
radical-radical/radical-molecular reaction pathways) are also written with their
rate coefficients along with the association energies. Details of the formation routes towards
the formation of interstellar pre-biotic molecules considered in the present paper are as follows:
\begin {figure}
\vskip 0.6cm
\centering{
\includegraphics[height=6cm,width=7cm]{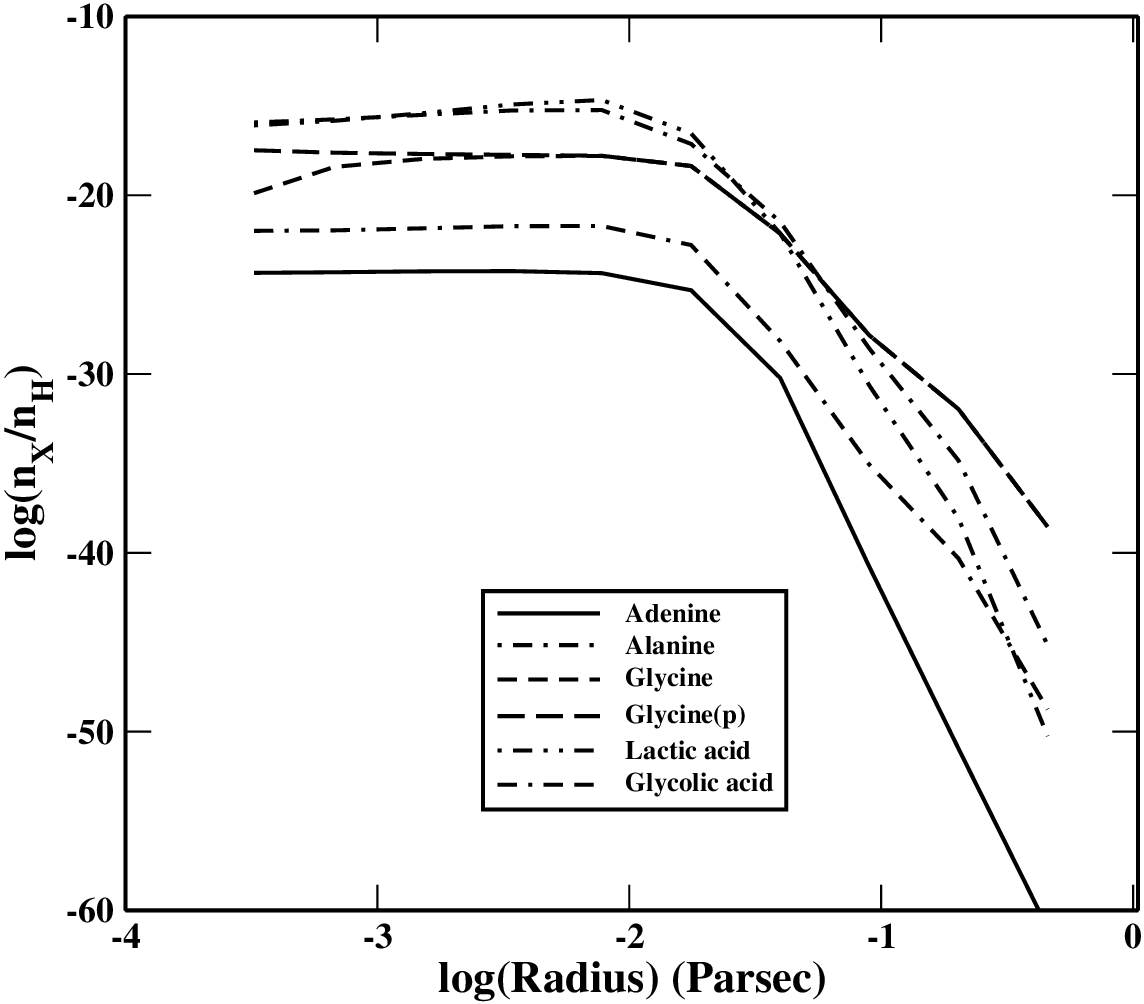}}
\caption{Final abundances of several pre-biotic molecules along different radial distances of a collapsing cloud
}
\label{fig-1}
\end {figure}

\subsubsection{Formation of adenine}
Adenine is an important constituent of the DNA molecule and its significant
production may provide an important clue into the problem of origin of life on
a planets like ours. Due to this reason, it is essential to carry out careful analysis
on the reaction rates during the adenine formation. Here we used the formation of
adenine via neutral-neutral and other variant reaction types by density functional calculations.
In the neutral-neutral reaction pathways, rate coefficient for the gas phase reaction
HCN+HCN$\rightarrow$CH(NH)CN is calculated to be 8.38$\times$10$^{-20}$ cm$^3$ s$^{-1}$.
This rate is very small since two neutral HCN molecule combine together by radiative association.
In the next step, CH(NH)CN is combining with the HCN to form NH$_2$CH(CN)$_2$ and the calculated
rate coefficient is 3.43$\times$10$^{-12}$ cm$^3$ s$^{-1}$. Similarly, NH$_2$CH(CN)$_2$ is
combining with the HCN to form NH$_2$(CN)C=C(CN)NH$_2$ and the rate coefficient is
3.30$\times$10$^{-15}$ cm$^3$ s$^{-1}$ and finally NH$_2$(CN)C=C(CN)NH$_2$ is combining with
the HCN to form adenine (C$_5$H$_5$N$_5$) and the rate coefficient is 3.99$\times$10$^{-10}$ cm$^3$ s$^{-1}$. 

Through the radical-radical/radical-molecular reaction pathways, adenine is formed in the interstellar
space through a barrierless exothermic reaction between hydrogen cyanide (HCN) and
cyanocarbene (HCCN). HCN is highly abundant around the star forming region and the
existence of HCCN in the ISM was reported by Guelien \& Cernicharo (1991).
Formation of HCCN on the grain surface was already considered by Hasegawa, Herbst \& Leung (1992).
In our gas phase chemical
network, we have considered this reaction between H and C$_2$N and the rate for the formation of
HCCN (2.67 $\times 10^{-14}$ cm$^3$sec$^{-1}$) has been calculated by following Bates (1983).
Following Park \& Lee (2002), we have considered the association reaction between HC and CN for the
production of HCCN, which is a barrierless and very exothermic in nature.
According to them this reaction might be considered as an efficient means
for the production of HCCN in the interstellar space. Here, the rate of formation
has been calculated by following Bates (1983).
Following Cherchneff, Glassgold \& Mamon (1993), we have incorporated all possible destruction reaction pathways
for the gas phase HCCN. According to them, the gas phase HCCN could be destroyed very efficiently by the
photo-dissociation reaction, having estimated rate coefficient 1.7 $\times$ 10$^{-9}$ sec$^{-1}$.

HCCN and HCN are both highly reactive carbine structures. The reaction between them leads to the formation of
1,2 dihydro imidazole (Molecule 1) with the rate coefficient 2.13$\times$10$^{-17}$ cm$^3$ s$^{-1}$.
In an exothermic addition reaction 1, 2 dihydro imidazole combines with a
hydrogen atom to form 2,3 dihydro-1H-imidazole radical (Molecule 2) with the rate coefficient
7.96$\times$10$^{-9}$ cm$^3$ s$^{-1}$. Existence of NH$_2$CN and CN radical
in the interstellar clouds was first reported by Fuente et al., (2005). 
Till then, these species were observed in both
diffuse and dense clouds by Liszt \& Lucas (2001). The reaction between the 2, 3 dihydro-1H-imidazole
radical and NH$_2$CN leads to the formation of 4 carboxaimidine-1H-imidazole (Molecule 3) radical with the
rate coefficient 6.24$\times$10$^{-12}$ cm$^3$ s$^{-1}$. Cyanide radical is very
abundant in the interstellar space (Fuente et al., 2005). The addition of CN radical with the
4 carboxaimidine-1H-imidazole radical leads to the formation of 2, 4 dihydro-3H-purine-6 amine (Molecule 4)
with the rate coefficient 1.80$\times$10$^{-9}$ cm$^3$ s$^{-1}$. In the next step, 2, 4
dihydro-3H-purine-6 amine combine with a hydrogen atom to form 6 amino-3H-purine (Molecule 5) with the rate
coefficient 8.71$\times$10$^{-9}$ cm$^3$ s$^{-1}$. At the end, the reaction between
the cyanide radical and Molecule 5 leads to the formation of adenine with the release of one HNC/HCN.
Calculated rate coefficient of this reaction is 1.89$\times$10$^{-9}$/1.91$\times$10$^{-9}$ cm$^3$ s$^{-1}$.
On the basis of results of the quantum chemical calculations presented above, it is evident that the
adenine formation by the radical-radical/radical-molecular reactions are more favourable than the neutral-neutral reactions
in terms of the rate coefficients. It would be of great interest to learn whether
any significant adenine is formed during the collapsing phase or not
and to compare between the different pathways for its formation.
It is possible that some of these reactions may take place on the icy grains also, but
that is outside the scope of the present paper.

\begin {figure}
\vskip 0.6cm
\centering{
\includegraphics[height=6cm,width=7cm]{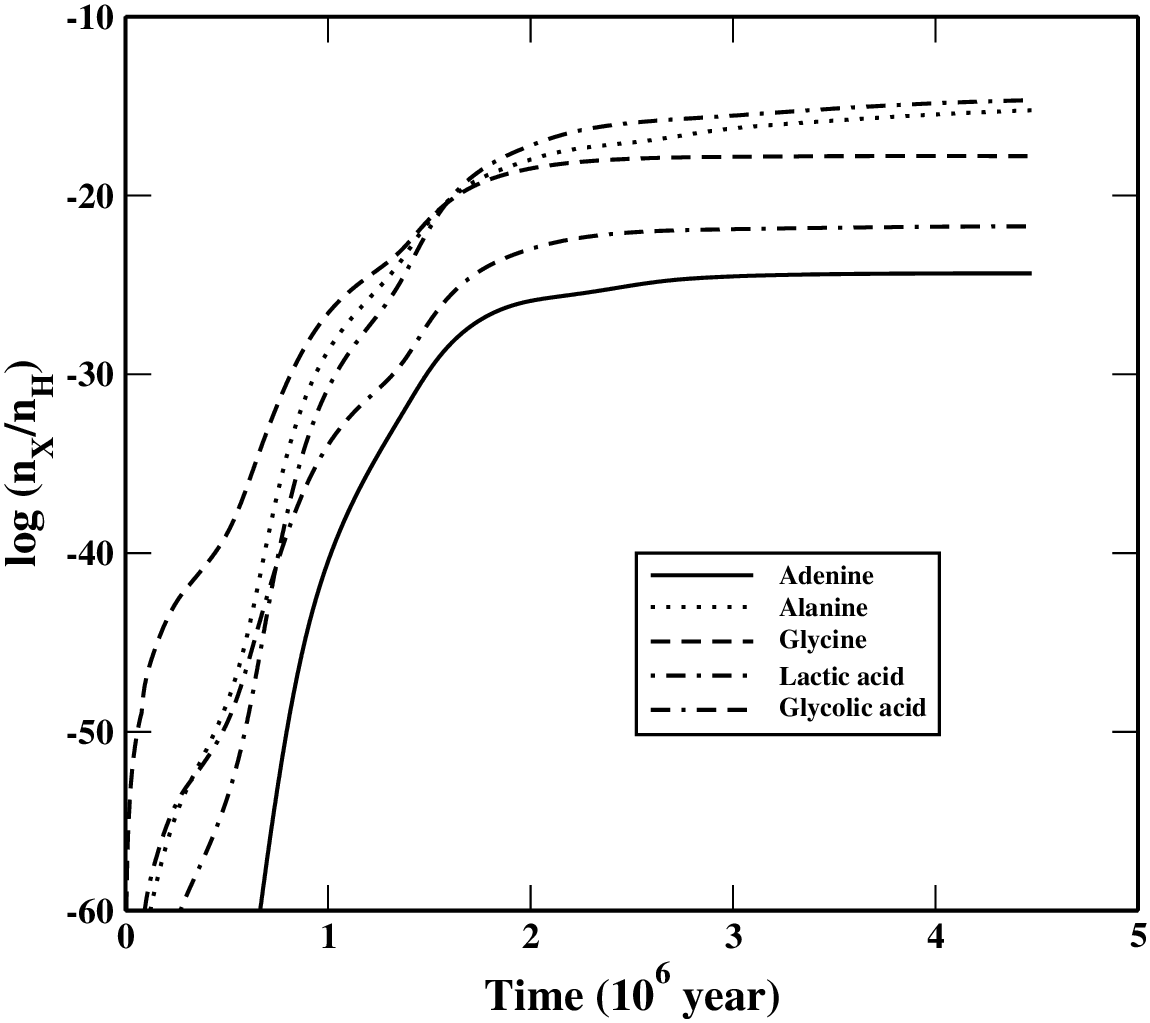}}
\caption{Time evolution of the interstellar pre-biotic molecules around an intermediate shell of a collapsing cloud.
}
\label{fig-1}
\end {figure}

\subsubsection{Formation of glycine}
Glycine (C$_2$H$_5$NO$_2$) is the smallest among the 20 amino acids commonly found in the proteins and also the only
pre-biotic molecule which was claimed to be observed in the ISM (Kuan et al., 2003), though there were a
number of unsuccessful attempts to confirm this observation (Snyder et al 2005). Guelin et al., (2008),
derived a 3 $\sigma$ upper limit on the column density of glycine of 1$\times 10^{15}$
cm$^{-2}$ per $2^{\verb+"+}$  $\times$ $3^{\verb+"+}$ beam in the Orion Hot Core and Compact Ridge.
Here we consider the formation of glycine via neutral-neutral
as well as radical-radical/radical-molecular pathways. Neutral-neutral pathways were taken from CC00ab and
radical-radical/radical-molecular pathways are taken from Woon (2002). Woon (2002) studied the formation of 
these molecules on UV irradiated ices. Here in our present study, we assume that these reactions
are also feasible in the gas phase.
In the case of the neutral-neutral pathways, H$_2$CO combines with the HCN to form
C$_2$H$_3$ON with the rate coefficient 1.23 $\times$10$^{-12}$ cm$^3$ s$^{-1}$ and produced
C$_2$H$_3$ON combines with H$_2$O to form interstellar glycine with
rate 2.68 $\times$10$^{-15}$ cm$^3$ s$^{-1}$.
In case of radical-radical/radical-molecular pathways, reactions between CH$_2$NH$_2$ and the HCN
leads to the formation of glycine. According to Woon (2002), CH$_2$NH$_2$ produce via
sequential hydrogenation of HCN and COOH originates from the reaction
between CO and OH with a rate coefficient 1.13 $\times 10^{-17}$ cm$^3$ s$^{-1}$,
where OH can be produced mainly through the photolysis of H$_2$O.
COOH then recombines with  CH$_2$NH$_2$ to form interstellar glycine
with a rate coefficient 1.82 $\times$10$^{-9}$ cm$^3$ s$^{-1}$.
According to Woon (2002), glycine formation by radical-radical/radical-molecular pathways most likely happen 
in such kind of interstellar ices, which could have experienced thermal shocks
or could have formed in comets that could have passed through warmer regions
of the solar system. In our model, we have assumed that these reactions are also 
feasible in the dense and cold gas as well, provided the cloud would have 
gone through the thermal shock regions. But if the cloud is dense enough
and interstellar radiation field does not able to penetrate deep inside, in that case
we could have disregard the radical-radical/radical-molecular pathways.

\subsubsection{Formation of alanine}
Alanine is an $\alpha$-amino acid with the chemical formula CH$_3$CH(NH$_2$)COOH (C$_3$H$_7$NO$_2$). The $\alpha$-carbon
atom of alanine is bound with a methyl group (-CH$_3$), making it one of the simplest $\alpha$-amino
acids. This group of alanine is non-reactive thus it is never directly involved
in protein function. Here we find out the rate coefficients for the formation of alanine via neutral-neutral
(CC00ab) and radical-radical/radical-molecular (Woon, 2002) reactions.
For the neutral-neutral reactions, CH$_3$CHO, HCN and water molecules are involved. All species are
reasonably abundant in the ISM. In the first step HCN and CH$_3$CHO react
together to form C$_3$H$_5$ON with the rate coefficient 1.84 $\times$10$^{-9}$ cm$^3$
s$^{-1}$ ((obtained from Eqn. 2, by using plorizibility 41.82 Bohr$^3$) and finally, C$_3$H$_5$ON
combines with the water molecule to form alanine (C$_3$H$_7$NO$_2$) with the rate coefficient
3.09 $\times$10$^{-14}$ cm$^3$ s$^{-1}$.
In the radical-radical/radical-molecular reaction pathways, NH$_2$CHCOOH and H are formed due to the fragmentation reaction
of glycine (NH$_2$CH$_2$COOH). As glycine is formed with enough internal energy by recombination
reaction of CH$_2$NH$_2$ and COOH, the fragmentation energy of the above reaction is readily supplied during  the
formation of glycine.
The rate coefficient of this reaction is 7.22 $\times$10$^{-10}$ cm$^3$ s$^{-1}$.
Finally, NH$_2$CHCOOH combines with CH$_3$ to form alanine (C$_3$H$_7$NO$_2$) with the rate
coefficient 2.30 x10$^{-11}$ cm$^3$ s$^{-1}$.
As like the glycine formation, 
Woon et al., (2002) discussed the production of alanine in the UV irradiated ice,
which are much warmer. So this pathways might not be useful for dense cloud.
The neutral neutral pathways as described
by CC00a could be very useful in this context, since these two reactions are exothermic in nature.

\subsubsection{Formation of glycolic and lactic acids}
Glycolic acid is the smallest $\alpha$-hydroxy acid while the lactic acid is a carboxylic acid. Both of
them are assigned to be the pre- biotic molecules. Due to 
the lack of any references in radical-radical/radical-molecular
reaction pathways for the formation of these important pre-biotic molecules, 
we only consider the neutral-neutral
pathways as used by CC00a though these are the endothermic reactions. 
The rate coefficient for the formation of lactic acid is
calculated to be 2.31 $\times$10$^{-9}$ cm$^3$ s$^{-1}$ { (obtained from Eqn. 2, by using
polarizability 98.22 Bohr $^3$)} and for glycolic acid it is 9.510 $\times$10$^{-15}$ cm$^3$ s$^{-1}$.

\subsection{Chemical evolution of the {pre-biotic} molecules}
\begin {figure}
\centering{
\includegraphics[height=7cm,width=9cm]{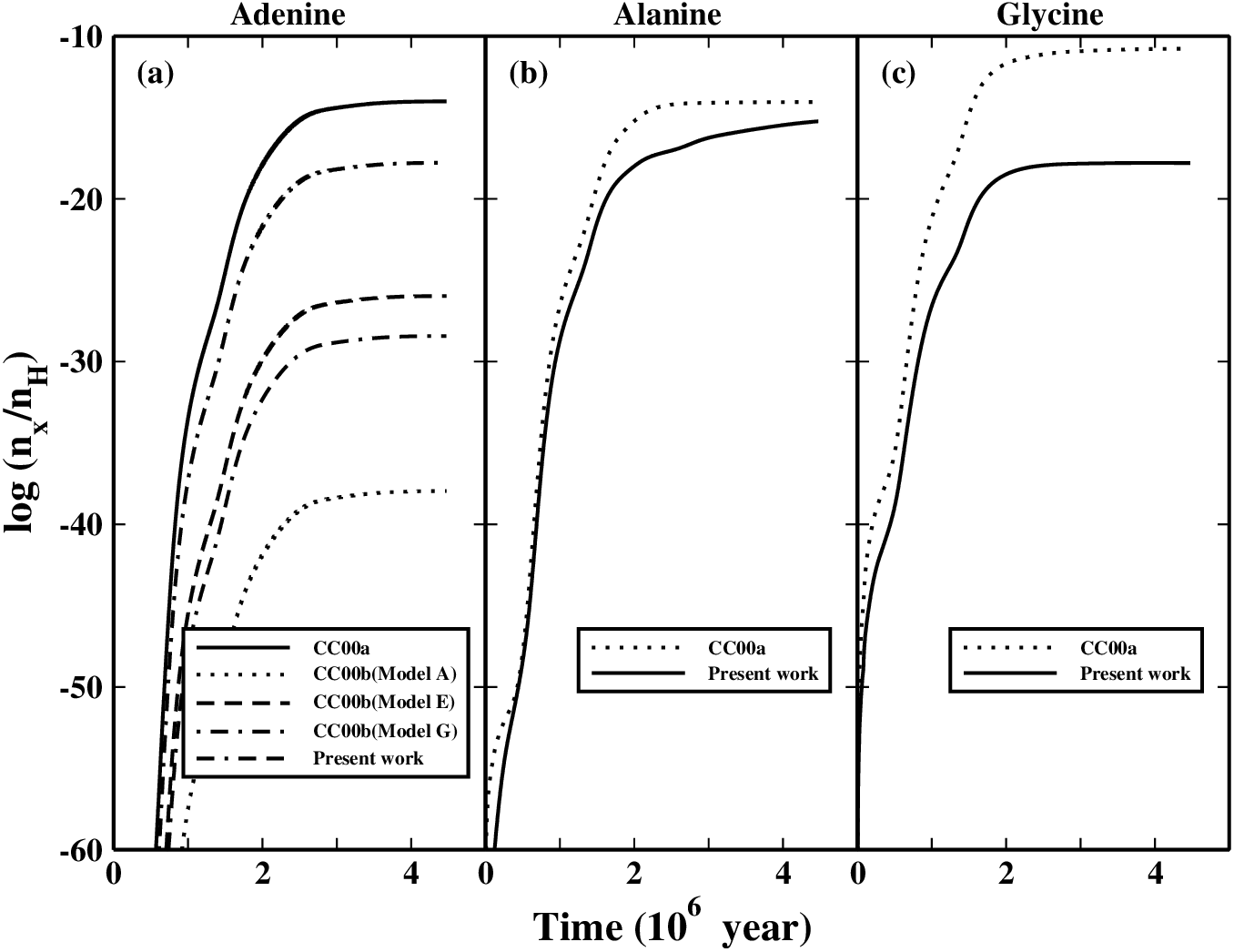}
}
\caption{Comparison of the calculated abundances of (a) adenine, (b) alanine and (c) glycine
using various prescriptions.}
\label{fig-1}
\end {figure}
We consider a large gas-grain network to study the chemical evolution of the biologically important interstellar
species. Separately a hydrodynamic simulation was carried out to obtain the physical parameters of a collapsing
cloud. Our simulation was carried out for a spherically symmetric isothermal (T=10K) interstellar cloud
for more than 4 million years. This is the typical lifetime of a molecular cloud (Das et al., 2010, 2011).
Though the calculated rate coefficients, which are calculated by using semi-empirical relation 
developed by Bates (1983), are applied for 30K, we assume that implementation of this rate coefficients 
into our present 10K model, would be within the error bar.

In Fig. 1, the time evolution of the number densities around different shells are shown.
As expected, deep inside the cloud the density reaches its maximum value. We took the density distribution of the
collapsing cloud as the input parameter of our chemical model to explore the interstellar medium. In Fig. 2, the variation of the
final abundances of the interstellar pre-biotic molecules are shown with the depth of the cloud.
In the `Y' axis, n$_X$ is denoted for the bio-species and n$_H$ is denoted for the total hydrogen nuclei in all forms.
It is evident from Fig. 2 that the abundances are peaking inside around 4$^{th}$ to 5$^{th}$ shell
(in case of glycine, it is peaking around the 1$^{st}$ shell). In case of glycine, we plot the peak abundance
also along with the final abundance. Glycine is decreasing due to the 
formation of C$_2$H$_4$O$_3$ (C$_2$H$_5$NO$_2$+ H$_2$O$\rightarrow$C$_2$H$_4$O$_3$+NH$_3$)
by the neutral-neutral reaction and formation of C$_2$H$_4$NO$_3$ by the 
fragmentation reaction (C$_2$H$_5$NO$_2$$\rightarrow$C$_2$H$_4$NO$_3$+H).
Peak abundance obtained during the simulation time scales are 
4.43$\times 10^{-25}$, 5.83$\times$10$^{-16}$,  1.57$ \times 10^{-18}$,
2.14 $\times 10^{-15}$, 1.89 $\times 10^{-22}$ for the interstellar adenine, alanine, glycine, lactic acid \& glycolic acid receptively.
If we have excluded the radical-radical/radical-molecular pathways for the production of glycine and alanine due to the reason
mentioned in the context of glycine and alanine, peak abundance obtained during the simulation time scales 
are 4.43$\times 10^{-25}$, 8.85$\times$10$^{-18}$, 1.96$ \times 10^{-17}$
, 2.35 $\times 10^{-16}$, 1.36 $\times 10^{-21}$ for the interstellar adenine, alanine, glycine, lactic acid \& glycolic acid receptively
around the 4$^{th}$ to 5$^{th}$ shell. Abundance of adenine remain unchanged since excluded pathways does not influence 
its production at all. Abundance of glycine and glycolic acid experiences a huge jump due to this exclusion. Since in the
pathways of Woon (2002), glycine readily dissociates into NH$_2$CHCOOH and H but now glycine could only channelized to form glycolic
acid by reacting with water molecule.

Since the abundances of most of the {pre-biotic} molecules are peaking at intermediate shells, in Fig. 3,
the time evolution of the pre-biotic molecules are shown for the 5$^{th}$ shell only. We started our simulation by
assuming the initial elemental abundances same as mentioned in Woodall et al. (2007).
We assumed that initially all the hydrogens were in the atomic stage.
As the time evolves, all the species start to become complex enough and initiate to
form some pre-biotic molecules. In Fig. 3, we plot the `X' axis in normal scale, whereas `Y' 
axis is shown in log scale. This is  because, at the beginning,
formation of pre-biotic molecules were highly hindered due to not having enough abundance of 
its required precursor species. Once the complex species
which are required to progress the chemical network start to form, the abundance of 
the pre-biotic molecules also start to grow. During
the simulation time scale, a near-steady state is attained by
the stable complex molecules. The cause of the decrease in glycine abundance has been mentioned 
in the context of Fig. 2 above.
The first attempt to quantify the interstellar adenine was made by CC00a who
proposed HCN addition reactions for its production. 
The rate coefficient for the successive HCN addition reactions were
assumed to be 10$^{-10}$ cm$^{3}$ s$^{-1}$.
Implementing the outcome of the quantum chemical calculation into our hydro-chemical model,
it is verified that the radical-radical/radical-molecular interaction was found to be more suitable
for the production of adenine in the ISM.
In CC00b, the rate coefficients were parameterized by assuming that as the 
HCN addition reaction goes on, the size of the molecule increases and thus
the rate coefficient also increases proportional to the area in each of the 
intermediate steps of the adenine formation
reaction. They assumed that the reaction rate constant varies as follows;
\begin{equation}
\alpha_i =f^{i-1} \alpha_{Ad},
\end{equation}
where, i is the number of steps involved (in CC00ab adenine formation require 4 steps), 
$\alpha_{Ad}$ and f were used to parameterized the reaction rates.
By parameterizing the reaction rate constant, they ran 7 models, namely; 
Model A ($\alpha_{Ad}=10^{-16}$ and f=1), 
Model B ($\alpha_{Ad}=10^{-14}$ and f=1), Model C ($\alpha_{Ad}=10^{-12}$ and f=1), 
Model D ($\alpha_{Ad}=10^{-10}$ and f=1, same as CC00a), Model E ($\alpha_{Ad}=10^{-16}$ and f=100), 
Model F ($\alpha_{Ad}=10^{-14}$ and f=10), Model G ($\alpha_{Ad}=10^{-12}$ and f=5).

According to CC00a (i.e., Model D of CC00b),
the abundance of adenine was predicted to be 6.35 $\times 10^{-11}$ when the 
effective rate, 10$^{-10}$ cm$^{3}$ s$^{-1}$ was used. In the
present simulation, even with the same rate constant as in CC00a, we find that the abundance is much lower at
9.84$\times 10^{-15}$. In CC00b, the adenine abundance was calculated
to be 1.36 $\times 10^{-34}$, 1.2$\times 10^{-22}$ and 1.8$\times 10^{-14}$ 
for Model A, E and G respectively. However,
for the same models and assuming the
same rate coefficients, our model calculates the adenine 
abundances 1.1$\times 10^{-38}$, 1.1 $\times 10^{-26}$, 1.65 $\times 10^{-18}$  respectively.
These numbers are several magnitudes lower. There are mainly two reasons 
for this differences.
First, we assume a different density distribution which were directly
arising out of a numerical simulation.  Most importantly, we extensively consider
here the gas-grain interaction simultaneously. We use
a larger network of grain chemistry along with a larger gas phase network.
Due to the presence of grains in our code, the gas phase HCN accrete onto the grain and as a result,
the gas phase HCN decreases. This in turn, decreases the production of adenine by a few orders of magnitudes.
Production of adenine by assuming our present quantum chemical calculations are also plotted 
in Fig. 4a (4.43$\times 10^{-25}$).
As a passing remark, we note that the reaction rate coefficients considered in Model E of
CC00b appears to be quite close to the actual adenine abundances reported in Fig. 4a.
In Fig. 4b and Fig. 4c, a comparison by assuming the parameters according to CC00a and
our present work have been carried out for the alanine and glycine respectively.
As most of the pre-biotic molecules having peak around the intermediate region, here also
we use the density distribution according to the 5$^{th}$ shell (Fig. 1) only.
From Table 1, it is clear that for some of the reactions, estimated rate coefficients
in CC00a, were few orders of magnitude higher in case of neutral-neutral reaction pathways and since neutral species
are very much abundant in comparison to the radicals for the formation of alanine/glycine, abundances obtained by
CC00a was much higher as compared to that obtained by us.
It is to be noted that throughout the lifetime of the collapsing cloud considered here, the production of adenine
is dominated by the radical-radical/radical-molecular reaction. This is because the reaction rates
involved in the 1$^{st}$ step of the radical-radical/radical-molecular pathways are 3 orders of magnitude higher than
the 1$^{st}$ step of the neutral-neutral pathways.

In case of adenine, NH$_2$(CN)C=C(CN)NH$_2$ (peak gas phase abundance 2.68 $\times 10^{-28}$)
and HCN (peak gas phase abundance 1.87 $\times 10^{-9}$)
recombine to form C$_5$H$_5$N$_5$ with a reaction rate of 3.99 $\times 10^{-10}$ cm$^3$ s$^{-1}$
in the last step of the neutral-neutral reaction network.
Its formation via radical-radical/radical-molecular reaction requires 6 amino 3H-purine (Molecule 5) (having peak gas phase abundance of 2.63 $\times 10^{-26}$) and CN (having peak gas phase abundance of
7.23 $\times 10^{-9}$) with a rate coefficient of 1.89 $\times 10^{-9}$ cm$^3$ s$^{-1}$, when HCN is produced as
an immediate product and 1.91 $\times 10^{-9}$ cm$^3$ s$^{-1}$ in case of HNC as an immediate product.
In the case of glycine production, CH$_2$NH$_2$, having peak gas phase abundance of 7.09$\times 10^{-11}$ and COOH, having peak
gas phase abundance of 3.53$\times 10^{-13}$ involved into the radical-radical/radical-molecular pathways with a reaction rate
coefficient 1.82$\times 10^{-9}$ cm$^3$ s$^{-1}$. Neutral-neutral pathways require C$_2$H$_3$ON, having a peak gas phase abundance of 7.29$\times 10^{-13}$
and H$_2$O, having a peak abundance of $3.74\times 10^{-8}$ with the rate coefficient of 2.68 $\times 10^{-15}$ cm$^3$ s$^{-1}$.
Though the neutral species for the formation of interstellar glycine are very much abundant in the gas phase, formation route
via radical-radical/radical-molecular pathways contributes $\sim 100\%$ of the production. The reason behind is that the
comparatively high rate coefficient is involved during the formation of glycine via radical-radical/radical-molecular pathways.

Interstellar alanine production by the neutral-neutral pathways requires C$_3$H$_5$ON which has a peak gas phase abundance of 5.24 $\times 10^{-13}$
and highly abundant gas phase H$_2$O with a rate coefficient of 3.09 $\times 10^{-14}$ cm$^3$ s$^{-1}$.
Radical-radical pathways require NH$_2$CHCOOH, having
a peak gas phase abundance of 7.49 $\times 10^{-14}$ and CH$_3$, having the peak gas phase abundance of 1.84 $\times 10^{-9}$ with a
rate coefficient of 2.3 $\times 10^{-11}$ cm$^3$ s$^{-1}$. An interesting trend is noted for the alanine production in the ISM. Throughout the process, a competition is going on between the two formation pathways.
Sometimes neutral-neutral pathway dominates over the radical-radical/radical-molecular pathway and sometimes the other way around.
This depends on the abundances of the reactive species involved at that instant.
                                                                                   
\section {Concluding Remarks}

In this paper, we  explore the possibility of formation of various pre-biotic molecules during the
collapsing phase of a proto-star. Our main goal was to compute the abundances with
realistic reaction rates and with the grain chemistry taken into account.
We observed that not only the reaction rates were important, the presence or absence of the grains made a difference of
several orders of magnitude in the final abundances of most of the pre-biotic molecules.
Despite the high abundances of the neutral species, it is noted that
in most of the cases, the radical-radical/radical-molecular reactions are dominating over the
neutral-neutral reaction pathways. The reason behind this is that the neutral-neutral
reactions very often possess high activation barrier energies, whereas in most of the
radical-radical and radical-radical/radical-molecular reactions are found to be barrier less in nature.
In case of adenine formation, it appears that radical-radical/radical-molecular reaction is dominant
(Smith et al. 2001), though, the abundance of the neutral species are much higher.
In CC00b, the reaction cross-sections for the neutral neutral reaction of adenine was
computed by assuming empirically that the reaction cross-sections
increases with the size of the reactants and indeed, the final abundances of Model E of CC00b predicts more or less
similar results to what we have found here, though present rates are more accurate.
In case of adenine, throughout the time evolution, the production
is dominated by the radical-radical/radical-molecular pathways. 
If the reaction pathways adopted from Woon (2002) for the production of alanine and glycine 
are assumed to be feasible in cold dense cloud then in the case of glycine, we find that the 
radical-radical/radical-molecular interaction plays the major role and  
in the case of alanine, both the pathways contribute significantly.

In the last decade, the subject of computing abundances of pre-biotic molecules has improved
considerably. The work presented in Das et al. (2008a)
represents significant improvements over earlier works in taking care of grain chemistry in a
collapsing cloud. This, along with the  improvements of the hydro-code to include the self-gravitating
cloud give more realistic understanding of the evolution of the chemical processes.
In future, we will re-look at these results by including rotation in the collapsing cloud which includes shocks and outflows.

The earth is a generic planet around a generic star, and, the processes to form pre-biotic molecules we described
in the present paper are also generic. Thus, it is likely that every collapsing cloud would produce these molecules
during the collapse phase. It is possible that during the planet formation stage the temperature went up sufficiently to
destroy these molecules. As was speculated in CC00a, the residuals of a planetary disk, such as comets could have
preserved these molecules and showered the planet much latter with these molecules
and thus providing the seeds of life in the planets inside the
so-called habitable zone. These are realistic speculations and works are under way to quantify them.

\section{Acknowledgments}
We acknowledges Suman Kundu of Ramakrishna Mission Residential College (Narendrapur)
for his helpful suggestions. L. Majumdar is grateful to DST for the financial support through a project (Grant No. SR/S2/HEP-40/2008)
and Ankan Das wants to thank ISRO respond project (Grant No. ISRO/RES/2/372/11-12).

\end{document}